# Eu-assisted enhancement of photoresponse in MBE-grown CdO/Si photodetectors

*Igor Perlikowski[1]\*, Eunika Zielony[1], Abinash Adhikari[2], Rafał Jakieła[2], Sergij Chusnutdinow[2], Ewa Popko[1], Ewa Przeździecka[2]*

[1] Department of Experimental Physics, Wroclaw University of Science and Technology, Wybrzeze Wyspianskiego 27, 50-370 Wroclaw, Poland

[2] Institute of Physics, Polish Academy of Sciences, Al. Lotnikow 32/46, 02-668 Warsaw, Poland

**Corresponding Author**

\* Igor Perlikowski – Department of Experimental Physics, Faculty of Fundamental Problems of Technology, Wroclaw University of Science and Technology, Wybrzeze Wyspianskiego 27, 50-370 Wroclaw, Poland; orcid.org/0000-0003-1724-942X; phone number: +48 71 320 2642; email: igor.perlikowski@pwr.edu.pl

**Abstract**

Doping cadmium oxide with rare earth (RE) elements is a way to control the band gap and enhance carrier concentration and mobility. This work presents how one of REs, europium, impacts performance of CdO/Si diode. The samples were grown using plasma-assisted molecular beam epitaxy. Doping level was modified by changing the temperature of the effusion cell with Eu and therefore flux of Eu particles. Different dopant concentrations were confirmed by secondary ion mass spectrometry. Atomic force microscopy images revealed a grain-like surface structure of the samples with grain size increasing after rapid thermal processing (RTP). Raman spectroscopy showed that introducing Eu changes vibrational properties of CdO through intraionic anharmonicity reduction. Kelvin probe method revealed upward band bending caused by oxygen adsorption during RTP. Electrical measurements confirmed that rectifying junctions were manufactured and that they are able to produce photocurrent in the spectral range of 450-1150 nm without external voltage bias. Introducing Eu into CdO was found to increase e.g. rectifying factor and responsivity. The results show that doping CdO with Eu is a way to enhance performance of the presented zero-power-consumption photodetectors, making it a promising material for future applications in optoelectronics.

**Keywords**

Rare-earth doped thin films, electrical properties, Raman spectroscopy, molecular beam epitaxy, CdO/Si photodetector

**1. Introduction**

Cadmium oxide (CdO) belongs to a wide group of transparent conductive oxides (TCO) with its direct bandgap of 2.18 eV at room temperature (RT) [1], along with two indirect, narrower ones [2]. Its electron mobility can reach over 300 cm$^2$/Vs [3], and an exceptional carrier concentration can be up to $10^{20}$ cm$^{-3}$ [4]. Pure and doped CdO has gained popularity for applications such as solar cells [5,6], photodetectors [7,8], and p-n junctions [9]. Additionally, ternary alloys containing CdO have recently been explored for optoelectronic applications [10]. To enhance device performance, CdO films can be doped with elements such as Mn [7,8], Fe [7], Sb, Sn, and Se [11]. Rare earth (RE) elements are particularly promising [12], as incorporating them provides an exquisite way to control the band gap [12–15] or increase carrier concentration and conductivity [12,14,16,17]. However, the addition of RE elements may also limit carrier mobility [12,14,17]. Several trials of successful fabrication of RE-doped CdO layers have been presented with various RE elements, including Sm, La [12,18], Nd [12], Eu [15], Ce, Pr [18]. When it comes to specific methods used for manufacturing these layers, magnetron sputtering [12], spray deposition [14,18], sol-gel spin coating [15], vacuum evaporation [13], successive ionic layer adsorption and reaction [19], and spray pyrolysis [17] belong to the most common ones.

Recent years have brought extensive research on CdO films and structures containing CdO layers fabricated by plasma-assisted molecular beam epitaxy (PA-MBE) [3,20–24] and in this work we would like to continue exploring that branch of semiconductor engineering. The fabrication of undoped CdO layers by PA-MBE has been thoroughly described by Adhikari et al. [3]. However, there is a lack of data on doped CdO, particularly with RE elements, grown by this method. The scientific literature primarily reports results for ion-implanted CdO films. Contrary to ion implantation, in situ doping does not cause damage to the crystal lattice. Thus, this study investigates CdO-based structures with in situ introduced europium as the RE element in the samples. Two valency states of Eu ions are most common: $Eu^{2+}$ and $Eu^{3+}$. In CdO they replace $Cd^{2+}$ ions. Especially $Eu^{3+}$ ions are beneficial, as they introduce surplus electrons leading to enhanced electrical conductivity [13,25]. Dakhel [13] has already demonstrated that Eu ions can significantly narrow direct band gap of CdO from 2.25 to 1.44 eV for 1.1% Eu content, as an effect of crystalline potential changes related to $Eu^{3+}$ incorporation. For 0.8% Eu concentrations it was managed to increase carrier concentration over 10 times, electron mobility improved 3.5 times, and resistivity decreased by nearly 40 times. However, the introduction of Eu in the structure can cause both optical band gap narrowing and widening. The latter occurrence is explained by the Burnstein-Moss effect [13,25,26]. Herein, we define the optical band gap as the energy range between the occupied states in the valence band and the first empty state in the conduction band. Functioning photodiodes based on CdO:Eu have been presented by Ravikumar et al. [17]. In this case, 3% Eu doping enhances photoresponsivity by 30%. Other examples of CdO:Eu-based electrical devices are not to be found in the literature. When it comes to undoped CdO/Si heterojunctions, published reports primarily discuss these structures in the context of solar cells or photodetectors. However, they often demonstrate low efficiencies and poor rectifying properties, implying that fabricating high-quality CdO/Si diodes remains a challenge for technologists. Hence, detailed electrical measurements of CdO:Eu/Si devices and thorough analysis of the collected data are needed to understand the current transport mechanisms in CdO:Eu/Si, optimize the growth process and discover the full potential of such junctions.

Given the mentioned gaps in semiconductor science, in this work we present photoactive p-n junctions consisting of n-CdO thin films doped in situ with Eu grown by PA-MBE on p-Si substrates. The influence of Eu content and rapid thermal processing (RTP) on structural and electrical properties was analyzed. Structural studies included secondary ion mass spectrometry (SIMS) to establish the Eu content, atomic force microscopy (AFM) to provide insight into surface morphology, and Raman spectroscopy to reveal changes in the vibrational properties of CdO caused by the presence of Eu ions and post-growth processing. The main part of the work is dedicated to electrical studies, starting with work function maps obtained using the Kelvin probe method. Then, basic electrical parameters were extracted from current-voltage characteristics. Finally, the potential applicability of CdO:Eu/Si diodes grown by PA-MBE in photodetectors and solar cells was confirmed by the measured responsivity spectra. They confirmed that our CdO/Si and CdO:Eu/Si junctions convert incident photons into current not only under reverse-biased conditions but also without any bias. This indicates that the samples can work as zero-power-consumption photodetectors, making this research promising for future applications in energy-saving microelectronics.

## 2. Experimental Details

Throughout this work, five structures based on CdO were investigated: an undoped CdO/Si junction (referred to as the CdO sample) and four CdO:Eu/Si diodes (cf. Figure 1a). Pure CdO and Eu-doped CdO layers were grown on p-type (100) Si substrates using a Riber Compact 21B plasma-assisted molecular beam epitaxy system. High-purity cadmium (6N) and europium (4N) were used as sources in effusion cells, with an RF-powered oxygen plasma used as the oxygen source. Before growth, the Si substrates were chemically etched with Buffered Oxide Etch (BOE) for 2 minutes, followed by wet and dry cleaning. The substrates were then annealed at 150°C for 1 hour in the load chamber. All layers were grown at 360°C, with a fixed Cd flux of $2.2 \times 10^{-7}$ Torr (effusion cell at 380°C). Oxygen flow was maintained at 3 sccm with constant RF power of 400 W. The structures differed in the Eu flux applied during the growth. It varied from $5.6 \times 10^{-9}$ Torr to $5.1 \times 10^{-9}$ Torr by adjusting the effusion cell temperature, $T_{Eu}$, from 300°C to 360°C. The Eu-doped structures were labeled Eu300, Eu320, Eu340, Eu360, corresponding to their respective $T_{Eu}$ values in °C. Selected fragments of the samples underwent rapid thermal processing (RTP) for 3 min at 900°C in an oxygen ($O_2$) atmosphere. To achieve that, AccuThermo AW 610 system from AllWin21 Corp. was used.

Raman spectra were collected using a HORIBA Jobin Yvon T64000 system configured for backscattering geometry and operating in a single subtractive mode. The system's spectrometer was set with slits totaled 0.1 mm, providing a spectral resolution of 0.5 $cm^{-1}$. A 532 nm semiconductor laser, with an output power of approximately 110 mW, was utilized to excite the samples. The laser beam, focused on the sample by a microscope lens, had a diameter of about 1 μm, and its power was reduced to 7 mW. The scattered light was detected by a CCD detector cooled with liquid nitrogen. Raman mode positions were extracted by fitting the spectra with Lorentz functions. Current-voltage (I-V) characteristics were measured at 300K using a Keithley 2601A I-V source meter, with stable temperature maintained by a LakeShore 325 Temperature Controller. Responsivity was measured with a Bentham PVE300 Photovoltaic Device Characterization System and a Zurich Instruments MFIA Impedance Analyzer. Light and dark I-V curves were recorded at 25°C using an I-V curve tracer and a PET Solar Simulator (#SS100AAA) operating at 1000 $W/m^2$ light intensity (AM1.5G). Atomic force microscopy (AFM) images were obtained using a Park Systems XE-70 system. Work function mapping was performed with an SKP5050 KP technology apparatus.

## 3. Results and Discussion

*3.1 Composition and Morphology: SIMS and AFM*

It is convenient to start the analysis from SIMS measurements to establish how the Eu content varies with the effusion cell temperature ($T_{Eu}$) in the structure. As shown in Figure 1b, for all samples except Eu340, the SIMS signals for Cd, O and Eu elements remain at the same level across sample's thickness, indicating a uniform distribution of Eu in these structures. However, for $T_{Eu}$=340°C, the Eu concentration significantly increases towards the Si substrate. Within the studied range of temperatures of the effusion cell with Eu, $T_{Eu}$, the average Eu concentration gradually increases from $2·10^{18}$ $cm^{-3}$ to $6.5·10^{18}$ $cm^{-3}$, as higher $T_{Eu}$ leads to an increased flux of the Eu. The SIMS results confirm the incorporation of Eu into the samples and demonstrate that manipulating with $T_{Eu}$ is an effective method for controlling the Eu concentration in CdO layers.

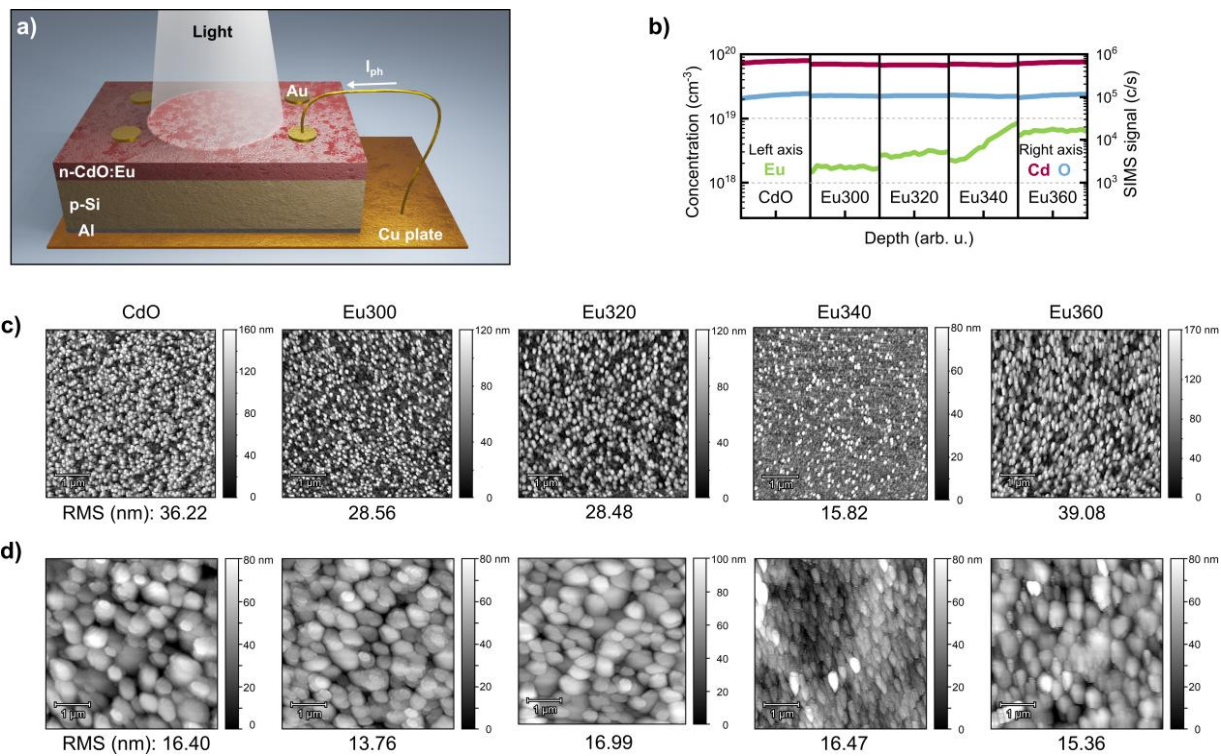

Figure 1. a) Schematic illustration of photocurrent generation in the investigated CdO:Eu/Si photodetector, b) SIMS depth profiles of Cd, O, Eu elements in pure CdO and Eu-doped CdO films grown on Si substrates, c) AFM images of 5 μm × 5 μm areas showing the grain-like surface morphology of as-grown CdO and Eu-doped CdO samples with calculated root mean square (RMS) surface roughness, d) AFM images of the samples annealed at 900°C.

Surface properties were studied using atomic force microscopy (AFM). Figure 1c compares the surface morphology of as-grown samples with and without Eu doping. Figure 1d presents the surface morphology of samples that underwent RTP at 900°C. These two groups of structures can be analyzed in terms of grain size and surface roughness. The images presented in Figure 1c reveal an average grain size of approximately 120-150 nm for all studied samples, indicating that the addition of Eu to the structure does not significantly affect the surface topography. This is likely due to the comparable ionic radii of $Cd^{2+}$ and $Eu^{3+}$ (0.097 i 0.095 nm, respectively [13]). Similar grain size of CdO thin films was obtained by Adhikari et al. for MBE-grown CdO layer on $Al_2O_3$ substrate [3]. In reference [25], it was shown that incorporating Eu decreases the grain size from 146 nm for pure CdO to 111 nm for 15% Eu in CdO.

From the images shown in Figure 1d, one may conclude that RTP significantly increased the grain size, from approximately 120-150 nm in the as-grown structures to over 300 nm in the majority of the grain-like shapes reported by us. This trend is consistent with other studies on CdO, which also report an increase in the grain size after RTP [27,28]. In contrast, surface roughness (calculated as RMS – root mean square) decreased after thermal processing from 28-40 nm for as-grown samples to ~15 nm after RTP. Similar roughness values for as-grown structures were observed in CdO and Eu-doped CdO films prepared with the sol-gel spin coating method [25]. Eu340 sample stands apart from the samples as, firstly, as-grown Eu340 has RMS of circa 16 nm. Secondly, RTP did not distinctly change RMS. This may be related to an inhomogeneous Eu doping distribution across the layer (cf. Figure 1b).

Considerations of grain size are crucial in the context of further analysis of the electrical properties of the samples. To determine whether the carrier concentration $n$ is homogeneous across the grains (and throughout the entire film), it is necessary to calculate the Debye length $L_D$ for CdO, which is expressed by the following equation [29,30]:

$$L_D = \left(\frac{\varepsilon\varepsilon_0 k_B T}{N_D q^2}\right)^{1/2}, \tag{1}$$

where $\varepsilon$ is the relative permittivity (20 for CdO [31]), $\varepsilon_0$ is the vacuum permittivity, $k_B$ is the Boltzmann constant, $T$ is the absolute temperature, $N_D$ is donor dopant concentration and $q$ is the elementary charge. Data on the carrier concentration in the investigated samples are not available. However, the literature indicates that for CdO, it typically ranges from $10^{18}$ to $10^{22}$ cm$^{-3}$, including pure and doped films, both as-grown and annealed samples [3,13,27,28,32–36]. An estimation of $L_D$ corresponding to this carrier concentration range is presented in Table 1. As shown, in all cases, $L_D$ is significantly smaller than half of the grain size, $l/2$, for each of the investigated samples. This suggests that carrier distribution is not homogeneous, and that grain boundaries contribute to an increased barrier height at the junction, which will be mentioned later in section 3.4.

Table 1. Debye length, $L_D$, of CdO calculated for various carrier concentrations.

| $N_D$ (cm$^{-3}$) | $L_D$ (nm) |
|---|---|
| $10^{17}$ | 16.92 |
| $10^{18}$ | 5.35 |
| $10^{19}$ | 1.69 |
| $10^{20}$ | 0.54 |
| $10^{21}$ | 0.17 |
| $10^{22}$ | 0.05 |

*3.2 Vibrational Properties from Raman Spectroscopy*

To examine the vibrational properties of the samples, Raman spectra were analyzed. Figure 2a illustrates the collected spectra for the as-grown structures. Identifying all the relevant Raman modes is essential for further research. In all the samples, a broad band can be noticed in the range of 200-450 cm$^{-1}$, which is typical for CdO. Since CdO is a rock salt compound, first-order Raman modes are forbidden [37]. However, it is noted that degradation of symmetry related to the non-stoichiometry of the structure may allow their observation [38], which is applicable here (compare Figure 1b). Therefore, the origin of the modes has not been clear over the years. In this work, the best fit of the data was achieved with four separate modes, marked with numbers from 1 to 4. Mode 1, which appears at ~260 cm$^{-1}$, can be identified as 2TA(L) – a second-order transverse acoustic (TA) mode from the L point of the Brillouin zone [39], TA+TO(L) – a combination of transverse acoustic and transverse optical (TO) modes [40], or the first-order TO mode [38,41]. Inclusion of mode 2 at ~275-281 cm$^{-1}$ is crucial in appropriate reflecting the Raman spectra. The TO mode is observed at RT in the range of 255-310 cm$^{-1}$ [38,41,42], which could explain this observation. Regarding mode 3 at 312-322 cm$^{-1}$, bands in this region have been recognized as second order modes [40] or, more specifically, as 2LA bands – second-order longitudinal acoustic modes [37,38,43]. Additionally, recent studies indicate that the observed Raman scattering in the considered range of Raman shift may be enhanced by charge density fluctuations [39]. Finally, mode 4 is thought to be a 2LA(Δ) [39] or LO(L) – a longitudinal optical – band [38]. From Figure 2, it is clear that the presence of the Eu dopant in CdO affects the shape of the spectrum. The maximum intensity shifts toward lower energies, which is not due to a shift of any single mode but rather a significant change in the relative intensities between modes 1 and 2. While mode 2 is twice as intense as mode 1 in pure CdO sample, these two bands are of comparable strength in Eu-doped structures. This observation differs from other work on Eu-doped CdO films. In the spectra presented by Ravikumar et al. [17], the addition of Eu to the structure results in an increased intensity of the 260 cm$^{-1}$ mode. Moreover, in their work, this mode is separated from the others, whereas in our work, all the considered Raman modes are blended into one broad band. However, the introduction of Sn into the CdO structure is known to decrease the intensity of the TO mode [42]. Similarly, doping of CdO with Eu increases the carrier concentration. As a result, Coulomb screening increases and intraionic anharmonicity is reduced, leading to the decay of the TO mode (referred to here as Raman mode no 2) [42]. The extracted positions of all observed modes are collected in Table 2.

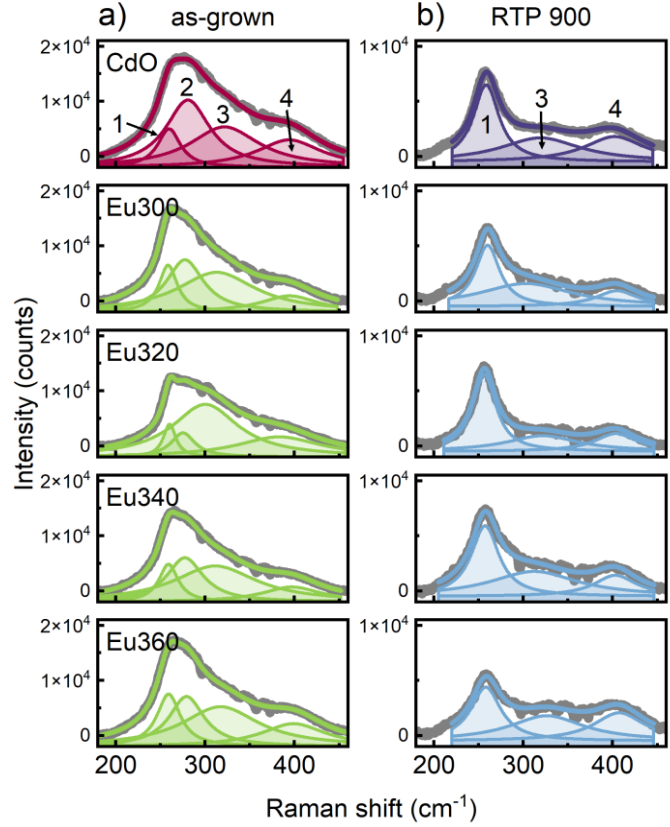

Figure 2. RT Raman spectra of a) as-grown and b) CdO and Eu-doped CdO films after RTP in 900°C.

Figure 2b presents the influence of rapid thermal processing (RTP) on the vibrational properties of the samples. RTP leads to a substantial decrease in the modes intensity. Notably, the shape of the spectra changes as well. After RTP, modes 1 and 4 become easily distinguishable, while mode 3 is relatively weaker. Mode 2 is evidently missing. This result is observed for all the structures. RTP in an oxygen atmosphere is known to decrease the number of oxygen vacancies in the material. Additionally, Cd and Eu can escape or migrate during RTP [44], which may be a significant factor explaining the behavior of the Raman bands. Any relevant differences between the pure CdO and Eu-doped CdO samples are no longer detectable.

Table 2. Peak positions of the observed Raman modes.

| Sample | Processing | Frequency of Raman Mode (cm$^{-1}$) | | | |
|---|---|---|---|---|---|
| | | 1 | 2 | 3 | 4 |
| CdO | as-grown | 260 | 281 | 322 | 394 |
| | RTP 900 | 258 | | 319 | 402 |
| Eu300 | as-grown | 259 | 278 | 313 | 395 |
| | RTP 900 | 260 | | 303 | 404 |
| Eu320 | as-grown | 260 | 275 | 300 | 383 |
| | RTP 900 | 256 | | 321 | 403 |
| Eu340 | as-grown | 259 | 278 | 312 | 397 |
| | RTP 900 | 257 | | 312 | 403 |
| Eu360 | as-grown | 259 | 279 | 317 | 399 |
| | RTP 900 | 258 | | 326 | 407 |

*3.3 Work Function Evolution in CdO:Eu Films*

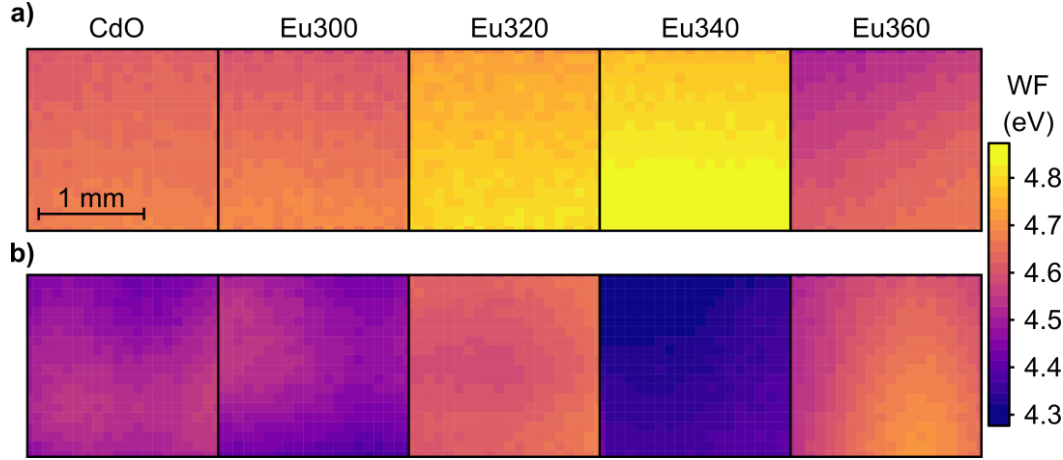

Figure 3. Images reflecting work function maps obtained from Kelvin probe measurements of the a) as-grown and b) annealed CdO/Si and CdO:Eu/Si samples.

Macro-scale (1.6 mm × 1.6 mm) work function ($WF$) maps were measured using the Kelvin probe method with a golden tip. To determine the $WF$ for each considered data point on the sample, the relation $WF = WF_{tip} - q \cdot V_{CPD}$ was used, where $WF_{tip}$ is the work function of the golden tip estimated at the beginning of the experiment, $q$ is the elementary charge, and $V_{CPD}$ is the measured contact potential difference. The results are presented in Figure 3. The images reveal visible inhomogeneity in the $WF$ distribution, with differences between extreme values reaching up to 0.1 eV. After RTP, the work function decreases in all samples except for Eu360. $WF$ can be expressed by [45]:

$$WF = (E_C - E_F)_{bulk} - qV_{bb} + \chi, \qquad (2)$$

where $E_C$ is the energy of the bottom of the conduction band, $E_F$ is the Fermi level. The difference between $E_C$ and $E_F$ concerns the bulk material's properties, excluding surface effects such as band bending due to surface charge and other unscreened charges. $V_{bb}$ describes band bending near surface and $\chi$ states for electron affinity. Doping affects the value of $E_C - E_F$ because the additional electrons introduced by the dopant shift $E_F$ away from the valence band. Doping is known to change $\chi$ as well [46]. As investigating these effects would require further research, the focus here is put on the factors influenced by RTP. Rapid thermal processing should not significantly influence $\chi$. It may slightly change $(E_C - E_F)_{bulk}$ due to minor modifications in the band gap [47]. However, RTP can affect the carrier concentration, $n$, by shifting the $E_F$ [28], which in turn changes the value of $(E_C - E_F)_{bulk}$. Comparisons of $n$ measured for as-grown vs. annealed samples are limited in the literature. Xie et al. [28] showed that for CdO films grown by radio frequency magnetron sputtering, RTP at 500°C under vacuum does not impact significantly $n$ in the case of pure CdO, but it does increase $n$ for CdO:Y films up to 9% Y content. Lysak et al. [44] investigated MBE-grown Eu-doped {ZnCdO/ZnO} superlattices and reported that RTP at 700 and 900°C in $O_2$ atmosphere may lead to a slight decrease in Eu concentration across samples, which could limit the dopant-related carrier concentration. In undoped CdO, defects are the main factors influencing the $n$ level, specifically oxygen vacancies ($V_O$) and cadmium interstitials [28]. Yu et al. [48] state that annealing sputter-deposited CdO in $O_2$ leads to a drop of $n$ due to $V_O$ reduction. However, this does not apply to oxygen-rich CdO. In the same work, the decrease in carrier concentration in In-doped CdO after annealing in $O_2$ is attributed to deactivation of dopants due to oxidation. Thus, based on the literature, it may be concluded that in our case, we could observe a reduction in $n$ after RTP, leading to a shift of $E_F$ toward valence band. However, this results in an increase in both $(E_C - E_F)_{bulk}$ component and $WF$, according to eq. (2), which contradicts our observations (cf. Figure 3). To explain the behavior of the work function after RTP, one must consider the band bending $V_{bb}$. Our observations suggest that annealing induces an upward bending of the electronic bands near the surface. The RTP was performed in $O_2$, which reduces the number of $V_O$ defects. Moreover, oxygen ions adsorbed on the surface contribute to increased band bending. A similar effect of oxygen adsorption following annealing in air has been observed in another oxide compound, $TiO_2$ [49]. Both surface states and defects influence $V_{bb}$ [50], making it a crucial factor for explaining the results. To summarize, we propose that RTP leads to increases in both the $(E_C - E_F)_{bulk}$ and $qV_{bb}$ components, which in turn affects the $WF$. While a general drop in $WF$ is observed, the changes in band banding dominate this behavior, except for the Eu360 sample with the highest Eu content, where no distinguishable differences were noted.

The investigated structures contain polycrystalline CdO thin films, so the measured CPD represents an average over a large number of grains. Both the individual grains and the band bending at the grain boundaries must be taken into account in the analysis.

*3.4 Electrical performance of CdO:Eu/Si junction*

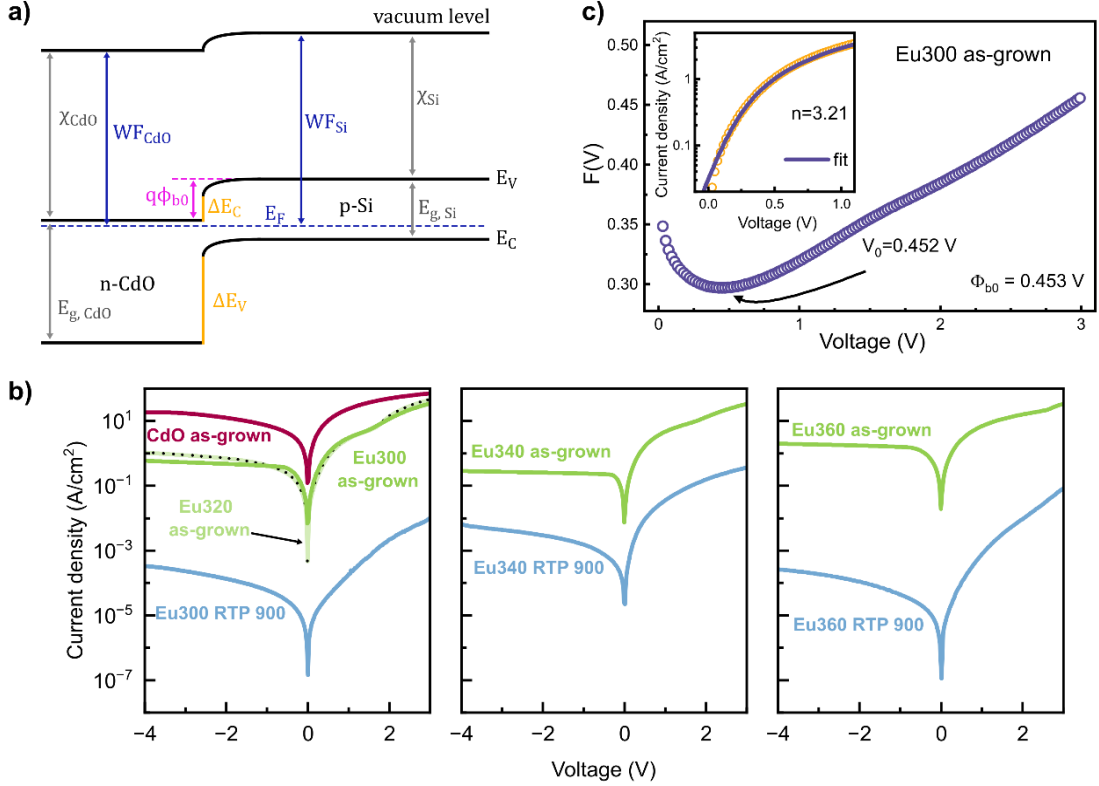

Figure 4. a) Schematic band diagram of the investigated CdO/Si heterojunction, b) current density-voltage characteristics of the CdO/Si and CdO:Eu/Si samples measured at 300K. Dots are added to the Eu300 as-grown curve to enhance readability, c) calculated F(V) plot for the Eu300 as-grown sample at 300K. The arrow points to the minimum of the F(V) function, which corresponds $V_0$. The inset shows the J-V curve fitted with eq. (2).

The examined structures are heterojunctions, as schematically illustrated in Figure 4a. To construct the band diagram, the following parameters were used: $E_{g,CdO}$ = 2.18 eV [1], $\chi_{CdO}$ = 4.51 eV [46] and $E_{g,Si}$ = 1.12 eV, $\chi_{Si}$ = 4.05 eV [51], which correspond to the energy gap and electron affinity of CdO and Si, respectively. Band offsets were determined as follows: $\Delta E_C$ = $\chi_{CdO}$-$\chi_{Si}$ = 0.46 eV, $\Delta E_V$ = $E_{g,CdO}$-$E_{g,Si}$+$\Delta E_C$ = 1.52 eV, where $WF_{CdO}$ and $WF_{Si}$ denote the work functions of CdO and Si, respectively. According to Figure 3, $WF_{CdO}$ is approximately 4.65 eV, suggesting that $E_F$ lays within the band gap of CdO. The zero-bias barrier height is marked as $\phi_{b0}$. $E_V$ represents the energy of the maximum of the valence band. To analyze the electrical properties of the junctions, current density-voltage (J-V) characteristics at 300K were measured, as shown in Figure 4b. The presence of the Eu dopant reduces leakage current and increases the rectifying factor. A similar trend is observed in the structures after the RTP process.

CdO itself has high carrier (electron) concentration, and doping with Eu further increases it. Hence, the carrier concentration on the n-side of the investigated junctions is approximately 2-3 orders of magintude higher than that of the p-Si substrate (~$2 \cdot 10^{17}$ cm$^{-3}$). As a result, these p-n heterojunctions may be analyzed analogously to metal-semiconductor junctions. Thus, the current flow through the junction is described by a formula [52,53]:

$$I = J \cdot A = I_0 \exp\left(\frac{q(V - IR_S)}{nk_BT}\right)\left[1 - \exp\left(-\frac{q(V - IR_S)}{k_BT}\right)\right], \quad (3)$$

where $J$ is the current density, $A$ is the contact area, $R_S$ is the series resistance, $n$ is the ideality factor, $T$ is the temperature of the measurement, and $I_0$ is the reverse saturation current (with $J_0$ being the corresponding current density) expressed by [52,53]:

$$I_0 = A \cdot J_0 = AA^*T^2 \exp\left(-\frac{q\phi_{b0}}{k_BT}\right). \quad (4)$$

In the formula (3), $A^*$ is the Richardson constant that for p-type Si equals to 32 A·cm$^{-2}$·K$^{-2}$ [52], and $\phi_{b0}$ is the zero-bias barrier height.

By fitting the J-V data in the low forward voltage bias region with eq. (3), the value of $n$ was extracted, as shown in the inset in Figure 4c. To determine $\phi_{b0}$, the Norde method was used. According to this method, the $F(V)$ function is calculated using the formula [52,53]:

$$F(V) = \frac{V}{\gamma} - \frac{k_BT}{q}\ln\left(\frac{I(V)}{AA^*T^2}\right), \quad (5)$$

where $\gamma$ is an arbitrarily chosen integer larger than $n$ (here, $\gamma=10$ was used). The obtained $F(V)$ function is presented in Figure 4c. The Norde method requires finding the voltage $V_0$ that corresponds to the minimum of $F(V)$ (compare with Figure 4c) [54]. $\phi_{b0}$ is then calculated as follows [52,53]:

$$\phi_{b0} = F(V_0) + \frac{V_0}{\gamma} - \frac{k_BT}{q}, \quad (6)$$

Table 3. Electrical parameters of the investigated junctions at 300K: rectifying factor $RF$ at ± 2 V, and zero-bias barrier height $\phi_{b0}$.

| Sample | Processing | RF at ±2 V | $n$ | $\phi_{b0}$ (V) |
| --- | --- | --- | --- | --- |
| CdO | as-grown | 4 | 1.7 | 0.38 |
| Eu300 | as-grown | 25 | 3.2 | 0.45 |
| | RTP 900 | 17 | 2.4 | 0.70 |
| Eu320 | as-grown | 21 | 3.9 | 0.47 |
| Eu340 | as-grown | 52 | 2.8 | 0.44 |
| | RTP 900 | 53 | 2.7 | 0.58 |
| Eu360 | as-grown | 9 | 2.7 | 0.42 |
| | RTP 900 | 87 | 4.4 | 0.72 |

All of the parameters extracted from the J-V data are collected in Table 3. The rectifying factor (RF) was calculated for ±2V. Each Eu-doped samples has a higher RF compared to the undoped one. At room temperature, single-digit values of RF are commonly recorded for CdO/Si junctions fabricated with other methods [11,17,55–58]. However, CdO/Si diodes with exceptional rectifying properties have been reported, with RF values reaching $10^6$ [59]. The highest RF value of 87 at 300K is observed for the Eu360 sample after RTP at 900°C. This result is significantly better than those reported for other Eu-doped CdO/Si diodes, where RF typically reaches around 10 in similar devices [17].

At 300K, the ideality factor $n$ for the CdO sample totals 1.7, indicating that the current flow through the junction is a combination of diffusion (where $n=1$) and recombination (where $n=2$) currents [60]. This is not the case for other samples, as $n$ exceeds 2. Such high values of $n$ can be attributed to other mechanisms, such as tunneling, shunt resistance (leakage current) or carrier trapping [60–62]. The high ideality factors observed in Eu-doped samples may result from increased defect-assisted tunneling or carrier trapping mechanisms introduced by the dopant. Furthermore, posssible spatial inhomogeneities at the junction interface could also lead to barrier height fluctuations and enhanced recombination, further contributing to the deviation from ideal diode behavior. A more detailed understanding would require complementary studies, such as temperature-dependent current–voltage analysis.

The zero bias-barrier height $\phi_{b0}$ for CdO reaches 0.38 V. Introducing Eu into the CdO enhances the barrier height by 10-24% in every investigated case. Additionally, RTP leads to a further increase in $\phi_{b0}$. Knowing that in all of the cases, the condition $\frac{l}{2} > L_D$ is fulfilled, the height of potential barrier $E_B$ at grain boundary is independent of the grain size and it is given by [63]

$$E_B = \frac{q^2 Q_t^2}{8\varepsilon\varepsilon_0 n}, \tag{7}$$

where $Q_t$ is trap density at grain boundaries. As we deduce that RTP cause reduction in $n$, the resultant increase in $E_B$ contributes to the observed rise in the measured $\phi_{b0}$.

*3.5 CdO:Eu films for zero-power-consumption photodetection*

To demonstrate the potential abilities of CdO:Eu/Si devices grown by PA-MBE for use as photodetectors, dark and light J-V curves were measured with Solar Simulator, under 1-sun illumination (AM1.5G), as presented in Figure 5. The entire surface of the samples was illuminated during the measurement. The photocurrent at 0 V appears very low. The explanation is as follows: at zero bias, carrier separation relies solely on the built-in electric field, which is insufficient to efficiently collect carriers generated in regions far from the contact. Under reverse bias, the stronger electric field enhances carrier collection across the entire illuminated area (about 5 × 10 mm$^2$, depending on the sample), resulting in a much higher photocurrent.

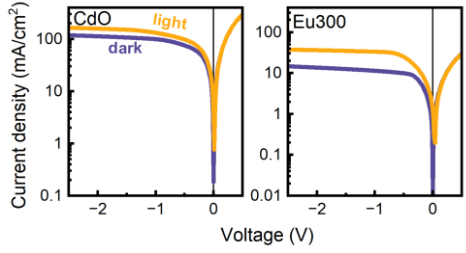

Figure 5. Comparison of dark and light J-V curves measured for as-grown reference and Eu-doped sample at 25°C. Light curves were obtained under 1-sun illumination (1000W/m², AM1.5G).

For efficient photocurrent generation across the entire surface of the structure at 0 V, the layout of the contacts deposited on the CdO:Eu layer needs to be optimized. Therefore, in the following measurements, in order to better demonstrate the potential of CdO:Eu/Si structures, local responsivity was measured by illuminating an area of approximately 1.7 mm² located directly adjacent to the deposited Au contact (cf. Figure 1a). Responsivity is defined as the difference between $I_{Light}$ current that flows through the junction under illumination and the $I_{Dark}$ current measured in the dark, divided by the incident light power ($P$). This relationship is expressed by the following [64]:

$$R = \frac{I_{Light} - I_{Dark}}{P} \ . \tag{8}$$

The results of spectral characteristics of $R$ are presented in Figure 6.

The junctions convert photons in the wavelength range of 450-1150 nm into photocurrent. As observed, Eu-doping generally enhances the performance of the devices. To fully explain the shape of the spectrum, several factors need to be considered: absorption capabilities of a certain layer, recombination processes that can affect the photocurrent potentially limiting the detector efficiency, and thin-film interference which can change the observed spectrum. By taking these factors into account, one can better interpret the responsivity spectrum and understand the photodetection performance of the CdO:Eu/Si devices.

The infrared boundary of responsivity can be attributed to the absorption edge of the Si substrate, which limits the detection of longer wavelengths. The high-energy limit, on the other hand, is likely due to recombination processes occurring near the surface of the CdO layer [57]. A slight shift of this high-energy limit towards shorter wavelengths is observed in the Eu-doped samples. This shift is particularly noticeable in Figure 6a, where the data for CdO and Eu-doped CdO are plotted together.

Introducing Eu dopant into CdO increases carrier concentration and enhances the Burstein-Moss effect, which results in the widening of the optical band gap of CdO, explaining the observed shift [12]. The signal oscillations seen in Figure 6a-c are attributed to CdO thin-film interference. Notably, the samples reacts to incident light without requiring an external voltage bias (cf. Figure 6a-c). This property allows them to be utilized as zero-power-consumption photodetectors. However, when a -2 V bias is applied (cf. Figure 6d-f), the collected signal increases by two orders of magnitude, as shown in the inset of Figure 6c. The explanation of this difference is as follows: $\phi_{b0}$ decreases when the junction is illuminated, facilitating photocurrent flow under reverse voltage bias[65–67]. Applying this voltage causes the built-in electric field within the junction to combine with the external electric field, allowing more photocarriers to contribute to the photocurrent as the depletion region expands. Moreover, the higher electric field provides better separation of electron-hole pairs generated by the incident light [66,68]. The Eu300 sample after RTP is not shown in Figure 6 as no photocurrent generation was observed. The observed degradation in diode performance following RTP – particularly in terms of current density (cf. Figure 4b) and photocurrent generation (cf. Figure 6b,c) – may be attributed to interdiffusion at the junction interface or to adverse changes in the structural and electronic properties of the CdO:Eu layer, leading to reduced carrier transport and weakened junction behavior. Additionally, thermally induced defect formation or an increased density of recombination centers at the interface could further contribute to the observed drop in device performance.

The differences in the level of responsivity between doped and undoped samples can be analyzed in the context of the structural properties of their surfaces. RMS calculated for as-grown Eu300, Eu320 and Eu340 is lower than that for the CdO sample (detailed results and analysis of surface properties will be published in another work). This suggests a lower level of light scattering, higher transmittance [12] and, as a result, a higher photocurrent measured with no voltage bias. The RMS of as-grown Eu360 sample is in turn slightly larger than that of CdO, which, by analogous reasoning, explains the weakest signal of the Eu-doped structures when $V_{bias}$ = 0 V. Moreover, the relatively large RMS roughness observed in the as-grown Eu360 sample may contribute to its reduced photocurrent under zero-bias conditions, as increased surface roughness can enhance surface recombination and reduce the effectiveness of the built-in electric field, highlighting the importance of surface quality in self-powered photodetectors. For RTP structures, another factor impacting photocurrent generation is the reduction in CdO transmittance caused by annealing. Such an occurrence was observed by Sing et al. for electrodeposited thin films [27].

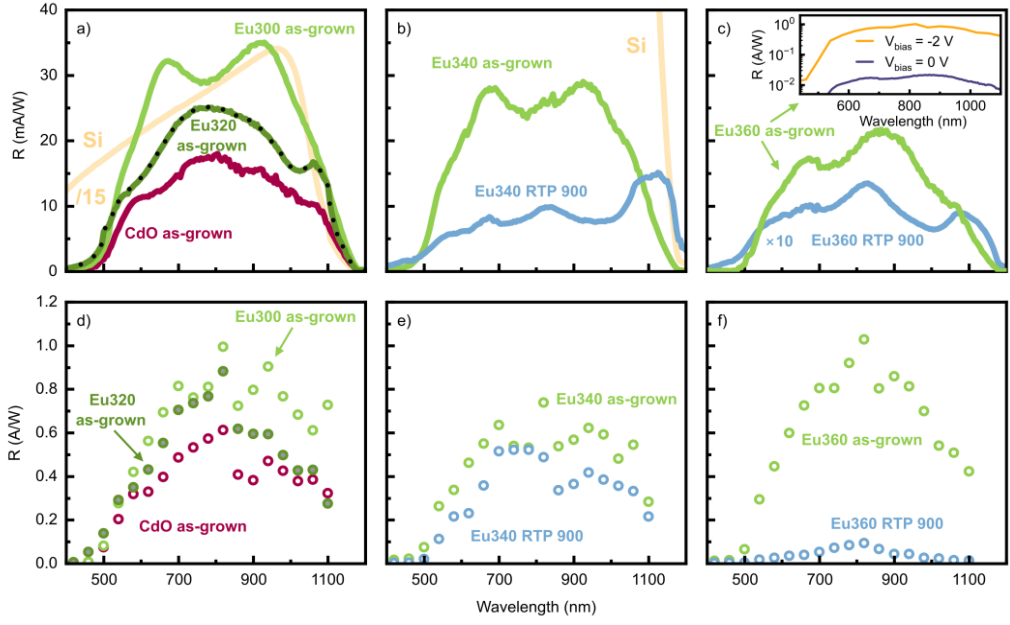

Figure 6. Responsivity ($R$) spectra measured with no voltage bias for a) as-grown CdO, Eu300, and Eu320 samples and a comparison between as-grown and RTP-treated structures, b) Eu340 and c) Eu360 samples. The responsivity spectrum of a standard Si detector is included for comparison. The Si spectrum was scaled in a). The inset in c) highlights the difference in signal levels collected with and without voltage bias for the as-grown Eu360 sample. Responsivity spectra measured with a voltage bias of -2 V for d) as-grown CdO, Eu300, and Eu320 samples, and comparison between as-grown and RTP-treated structures for e) Eu340 and f) Eu360 samples.

To further explore the potential of these structures as photodetectors operating without external voltage bias, the external quantum efficiency ($EQE$) was calculated based on the following formula [64]:

$$EQE = R \cdot \frac{hc}{q\lambda}, \qquad (9)$$

while specific detectivity $D^*$ is given by [64]:

$$D^* = \frac{R}{\sqrt{2q \cdot \frac{I_{Dark}}{A}}}, \qquad (10)$$

where $A$ is the active illuminated area of the device.

Comparison of spectral characteristics of $EQE$ and $D^*$ for the reference (undoped) CdO sample and the best-performing as-grown and annealed Eu-doped structures measured under 0 V bias are presented in Figure 7. Single digit values of $EQE$ can be perceived as low. However, while the literature presents devices achieving 20-80%, they require external voltage bias [64]. While $EQE$ of Eu340 RTP is the lowest, detectivity reaches over $10^{11}$ Jones, which is due to extremely low dark current at the level of $10^{-11}$ A. The detectivity of CdO/Si photodetectors in the literature is typically in the range of $10^8$ to $10^{12}$ Jones [55,64,69]. These results highlight the strong need for further research concerning the influence of RTP on interface quality and the reduction of dark current under zero-bias conditions.

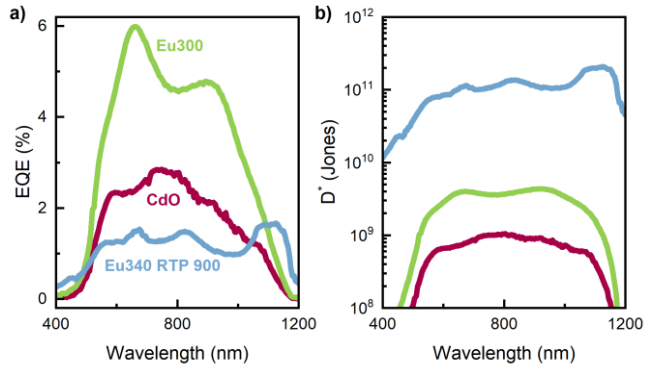

Figure 7. Comparison of spectral characteristics of a) external quantum efficiency ($EQE$) and b) specific detectivity ($D^*$) for the reference (undoped) CdO sample and the best-performing as-grown and annealed Eu-doped structures measured under 0 V bias.

**Conclusions**

This work reports in situ Eu-doped CdO layers grown by PA-MBE on Si substrate. SIMS data shows that obtained Eu concentration can be controlled by Eu effusion cell temperature and change from $10^{18}$ to $10^{19}$ cm$^{-3}$. Eu-doped samples can be distinguished from the undoped CdO one by comparing the respective Raman spectra, as Eu atoms present in the structure cause weakening of 278 cm$^{-1}$ Raman mode. Current density-voltage data demonstrates an increase of rectifying ratio and zero-bias barrier height, $\phi_{b0}$, as an effect of Eu doping. J-V curves under illumination demonstrates light-to-current conversion abilities of the samples. Responsivity spectra reveals that the structures are sensitive to photons from 450-1150 nm wavelength range. The samples can generate photocurrent without external voltage bias, making them a suitable choice for zero-power-consumption photodetectors. In such conditions, Eu doping clearly improves responsivity, for instance from over 15 mA/W (undoped sample) to almost 30 mA/W at 800 nm, when Eu concentration is at $2 \cdot 10^{18}$ cm$^{-3}$. This structure exhibited the best external quantum efficiency of 6% at 665 nm. Effects of rapid thermal processing at 900°C were investigated as well. Changes in work function after RTP are observed, attributed to a decrease in carrier concentration and upward band bending caused by oxygen surface adsorption during annealing. Enlargement of grain size caused by annealing was noted, from 110-150 nm for as-grown samples to over 300 nm after RTP. RTP caused growth of $\phi_{b0}$, which is assigned mainly to changes of the height of potential barriers at grain boundaries. Measured current densities in the diodes after thermal processing are significantly lower than in the as-grown samples. Only part of the structures after RTP were able to generate photocurrent. However, the highest detectivity at a level of $10^{11}$ Jones was achieved for an annealed Eu-doped structure mainly thanks to the low dark current.

The results proves that PA-MBE grown Eu-doped CdO is a suitable material for energy-saving optoelectronics. However, further optimalization of the growing process needs to be conducted, as the device performance is a complex interplay between roughness, dopant level and crystal quality.

**Author Contributions**

I.P. – conceptualization, formal analysis, investigation, writing – original draft, visualization; E.Z. – investigation, writing – review & editing, supervision; A.A. – resources; R.J. – investigation; S.C. – investigation; E.Po. – writing – review & editing; E.Pr. – writing – review & editing, validation, supervision

**Declaration of Competing Interest**

The authors declare that they have no known competing financial interests or personal relationships that could have appeared to influence the work reported in this paper.

**Data availability**

Data will be made available on request.

**Acknowledgements**

This work was supported by the Polish National Science Center (Grant Nos. 2021/41/N/ST5/00812 and 2021/41/B/ST5/00216).

**References**

[1] S.K. Vasheghani Farahani, V. Muñoz-Sanjosé, J. Zúñiga-Pérez, C.F. McConville, T.D. Veal, Temperature dependence of the direct bandgap and transport properties of CdO, Appl Phys Lett 102 (2013). https://doi.org/10.1063/1.4775691.

[2] Y. Dou, R.G. Egdell, D.S.L. Law, N.M. Harrison, B.G. Searle, An experimental and theoretical investigation of the electronic structure of CdO, Journal of Physics: Condensed Matter 10 (1998) 8447–8458. https://doi.org/10.1088/0953-8984/10/38/006.

[3] A. Adhikari, A. Wierzbicka, Z. Adamus, A. Lysak, P. Sybilski, D. Jarosz, E. Przezdziecka, Correlated carrier transport and optical phenomena in CdO layers grown by plasma-assisted molecular beam epitaxy technique, Thin Solid Films 780 (2023) 139963. https://doi.org/10.1016/j.tsf.2023.139963.

[4] K.M. Yu, M.A. Mayer, D.T. Speaks, H. He, R. Zhao, L. Hsu, S.S. Mao, E.E. Haller, W. Walukiewicz, Ideal transparent conductors for full spectrum photovoltaics, J Appl Phys 111 (2012). https://doi.org/10.1063/1.4729563.

[5] I.L.P. Raj, N. Chidhambaram, S. Saravanakumar, S. Sasikumar, S. Varadharajaperumal, D. Alagarasan, T. Alshahrani, Mohd. Shkir, S. AlFaify, A comprehensive study on effect of annealing on structural, morphological and optical properties of CdO and photodetection of heterojunction n-CdO/p-Si diode, Optik (Stuttg) 241 (2021) 166406. https://doi.org/10.1016/j.ijleo.2021.166406.

[6] A.S. Jasim, K.A. Aadim, J.M. Hussein, Preparation and fabrication of (Mg,Sn) doped CdO/PSi solar cell by laser induced plasma, IOP Conf Ser Mater Sci Eng 928 (2020) 072022. https://doi.org/10.1088/1757-899X/928/7/072022.

[7] B.R. Kumar, K.H. Prasad, K. Kasirajan, M. Karunakaran, V. Ganesh, Y. Bitla, S. AlFaify, I.S. Yahia, Enhancing the properties of CdO thin films by co-doping with Mn and Fe for photodetector applications, Sens Actuators A Phys 319 (2021) 112544. https://doi.org/10.1016/j.sna.2021.112544.

[8] K. Kasirajan, A.N.A. Anasthasiya, O.M. Aldossary, M. Ubaidullah, M. Karunakaran, Structural, morphological, optical and enhanced photodetection activities of CdO films: An effect of Mn doping, Sens Actuators A Phys 319 (2021) 112531. https://doi.org/10.1016/j.sna.2020.112531.

[9] K. Mohanraj, D. Balasubramanian, J. Chandrasekaran, A.C. Bose, Synthesis and characterizations of Ag-doped CdO nanoparticles for P-N junction diode application, Mater Sci Semicond Process 79 (2018) 74–91. https://doi.org/10.1016/J.MSSP.2018.02.006.

[10] M.A. Pietrzyk, A. Wierzbicka, E. Zielony, A. Pieniazek, R. Szymon, E. Placzek-Popko, Fundamental studies of ZnO nanowires with ZnCdO/ZnO multiple quantum wells grown for tunable light emitters, Sens Actuators A Phys 315 (2020) 112305. https://doi.org/10.1016/j.sna.2020.112305.

[11] H.I. Hussein, A.H. Shaban, I.H. Khudayer, Enhancements of p-Si/CdO Thin Films Solar Cells with doping (Sb, Sn, Se), Energy Procedia 157 (2019) 150–157. https://doi.org/10.1016/j.egypro.2018.11.175.

[12] P. Sakthivel, S. Asaithambi, M. Karuppaiah, S. Sheikfareed, R. Yuvakkumar, G. Ravi, Different rare earth (Sm, La, Nd) doped magnetron sputtered CdO thin films for optoelectronic applications, Journal of Materials Science: Materials in Electronics 30 (2019) 9999–10012. https://doi.org/10.1007/s10854-019-01342-9.

[13] A.A. Dakhel, Bandgap narrowing in CdO doped with europium, Opt Mater (Amst) 31 (2009) 691–695. https://doi.org/10.1016/j.optmat.2008.08.001.

[14] P. Velusamy, R.R. Babu, K. Ramamurthi, J. Viegas, E. Elangovan, Structural, microstructural, optical and electrical properties of spray deposited rare-earth metal (Sm) ions doped CdO thin films, Journal of Materials Science: Materials in Electronics 26 (2015) 4152–4164. https://doi.org/10.1007/s10854-015-2960-0.

[15] G. Turgut, G. Aksoy, D. İskenderoğlu, U. Turgut, S. Duman, The effect of Eu-loading on the some physical features of CdO, Ceram Int 44 (2018) 3921–3928. https://doi.org/10.1016/j.ceramint.2017.11.183.

[16] A.A. Dakhel, Optoelectronic properties of Eu- and H-codoped CdO films, Current Applied Physics 11 (2011) 11–15. https://doi.org/10.1016/j.cap.2010.06.003.

[17] M. Ravikumar, V. Ganesh, M. Shkir, R. Chandramohan, K.D. Arun Kumar, S. Valanarasu, A. Kathalingam, S. AlFaify, Fabrication of Eu doped CdO [Al/Eu-nCdO/p-Si/Al] photodiodes by perfume atomizer based spray technique for opto-electronic applications, J Mol Struct 1160 (2018) 311–318. https://doi.org/10.1016/j.molstruc.2018.01.095.

[18] V. Periyasamy, R.R. Babu, A. Ahmad, M.D. Albaqami, R.G. Alotabi, E. Elamurugu, Spray-Deposited Rare Earth Metal Ion ($Sm^{3+}$, $Ce^{3+}$, $Pr^{3+}$, $La^{3+}$)-Doped CdO Thin Films for Enhanced Formaldehyde Gas Sensing Characteristics, ACS Omega 7 (2022) 35191–35203. https://doi.org/10.1021/acsomega.2c04303.

[19] B. Şahin, Dual doping (Cu with rare-earth element Ce): An effective method to enhance the main physical properties of CdO films, Superlattices Microstruct 136 (2019) 106296. https://doi.org/10.1016/j.spmi.2019.106296.

[20] A. Lysak, E. Przeździecka, R. Jakiela, A. Reszka, B. Witkowski, Z. Khosravizadeh, A. Adhikari, J.M. Sajkowski, A. Kozanecki, Effect of rapid thermal annealing on short period $\{CdO/ZnO\}_m$ SLs grown on m-Al2O3, Mater Sci Semicond Process 142 (2022) 106493. https://doi.org/10.1016/j.mssp.2022.106493.

[21] E. Przeździecka, P. Strąk, A. Wierzbicka, A. Adhikari, A. Lysak, P. Sybilski, J.M. Sajkowski, A. Seweryn, A. Kozanecki, The Band-Gap Studies of Short-Period CdO/MgO Superlattices, Nanoscale Res Lett 16 (2021) 59. https://doi.org/10.1186/s11671-021-03517-y.

[22] A. Lysak, E. Przeździecka, A. Wierzbicka, P. Dłużewski, J. Sajkowski, K. Morawiec, A. Kozanecki, The Influence of the Growth Temperature on the Structural Properties of $\{CdO/ZnO\}_{30}$ Superlattices, Cryst Growth Des 23 (2023) 134–141. https://doi.org/10.1021/acs.cgd.2c00826.

[23] E. Przezdziecka, A. Wierzbicka, P. Dłuzewski, I. Sankowska, P. Sybilski, K. Morawiec, M.A. Pietrzyk, A. Kozanecki, Short-Period CdO/MgO Superlattices as Cubic CdMgO Quasi-Alloys, Cryst Growth Des 20 (2020) 5466–5472. https://doi.org/10.1021/acs.cgd.0c00678.

[24] E. Przeździecka, A. Wierzbicka, A. Lysak, P. Dłużewski, A. Adhikari, P. Sybilski, K. Morawiec, A. Kozanecki, Nanoscale Morphology of Short-Period {CdO/ZnO} Superlattices Grown by MBE, Cryst Growth Des 22 (2022) 1110–1115. https://doi.org/10.1021/acs.cgd.1c01065.

[25] A.A.M. Farag, M.I. Mohammed, V. Ganesh, H.E. Ali, A.M. Aboraia, Y. Khairy, H.H. Hegazy, V. Butova, A. V. Soldatov, H. Algarni, H.Y. Zahran, I.S. Yahia, Investigating the influence of Eu-doping on the structural and optical characterization of cadmium oxide thin films, Optik (Stuttg) 281 (2023) 170830. https://doi.org/10.1016/j.ijleo.2023.170830.

[26] S. Gahlawat, J. Singh, A.K. Yadav, P.P. Ingole, Exploring Burstein–Moss type effects in nickel doped hematite dendrite nanostructures for enhanced photo-electrochemical water splitting, Physical Chemistry Chemical Physics 21 (2019) 20463–20477. https://doi.org/10.1039/C9CP04132J.

[27] T. Singh, D.K. Pandya, R. Singh, Annealing studies on the structural and optical properties of electrodeposited CdO thin films, Mater Chem Phys 130 (2011) 1366–1371. https://doi.org/10.1016/j.matchemphys.2011.09.035.

[28] M. Xie, W. Zhu, K.M. Yu, Z. Zhu, G. Wang, Effects of doping and rapid thermal processing in Y doped CdO thin films, J Alloys Compd 776 (2019) 259–265. https://doi.org/10.1016/j.jallcom.2018.10.288.

[29] W.C. Johnson, P.T. Panousis, The influence of debye length on the C-V measurement of doping profiles, IEEE Trans Electron Devices 18 (1971) 965–973. https://doi.org/10.1109/T-ED.1971.17311.

[30] E. Przezdziecka, E. Guziewicz, D. Jarosz, D. Snigurenko, A. Sulich, P. Sybilski, R. Jakiela, W. Paszkowicz, Influence of oxygen-rich and zinc-rich conditions on donor and acceptor states and conductivity mechanism of ZnO films grown by ALD—Experimental studies, J Appl Phys 127 (2020). https://doi.org/10.1063/1.5120355.

[31] A. Crovetto, M.K. Huss-Hansen, O. Hansen, How the relative permittivity of solar cell materials influences solar cell performance, Solar Energy 149 (2017) 145–150. https://doi.org/10.1016/j.solener.2017.04.018.

[32] J. Santos-Cruz, G. Torres-Delgado, R. Castanedo-Perez, S. Jiménez-Sandoval, O. Jiménez-Sandoval, C.I. Zúñiga-Romero, J. Márquez Marín, O. Zelaya-Angel, Dependence of electrical and optical properties of sol–gel prepared undoped cadmium oxide thin films on annealing temperature, Thin Solid Films 493 (2005) 83–87. https://doi.org/10.1016/j.tsf.2005.07.237.

[33] M. Azizar Rahman, M.K.R. Khan, Effect of annealing temperature on structural, electrical and optical properties of spray pyrolytic nanocrystalline CdO thin films, Mater Sci Semicond Process 24 (2014) 26–33. https://doi.org/10.1016/j.mssp.2014.03.002.

[34] A.A. Dakhel, Effect of thermal annealing in different gas atmospheres on the structural, optical, and electrical properties of Li-doped CdO nanocrystalline films, Solid State Sci 13 (2011) 1000–1005. https://doi.org/10.1016/j.solidstatesciences.2011.02.002.

[35] N. Ueda, H. Maeda, H. Hosono, H. Kawazoe, Band-gap widening of CdO thin films, J Appl Phys 84 (1998) 6174–6177. https://doi.org/10.1063/1.368933.

[36] L. Wang, Y. Yang, S. Jin, T.J. Marks, MgO(100) template layer for CdO thin film growth: Strategies to enhance microstructural crystallinity and charge carrier mobility, Appl Phys Lett 88 (2006). https://doi.org/10.1063/1.2195093.

[37] R. Cuscó, J. Ibáñez, N. Domenech-Amador, L. Artús, J. Zúñiga-Pérez, V. Muñoz-Sanjosé, Raman scattering of cadmium oxide epilayers grown by metal-organic vapor phase epitaxy, J Appl Phys 107 (2010). https://doi.org/10.1063/1.3357377.

[38] Z. V. Popović, G. Stanišić, D. Stojanović, R. Kostić, Infrared and Raman Spectra of CdO, Physica Status Solidi (b) 165 (1991). https://doi.org/10.1002/pssb.2221650249.

[39] R. Cuscó, J. Yeste, V. Muñoz-Sanjosé, Temperature dependence of Raman scattering in CdO: Insights into phonon anharmonicity and plasmon excitations, Phys Rev B 107 (2023) 125204. https://doi.org/10.1103/PhysRevB.107.125204.

[40] R. Oliva, J. Ibáñez, L. Artús, R. Cuscó, J. Zúñiga-Pérez, V. Muñoz-Sanjosé, High-pressure Raman scattering of CdO thin films grown by metal-organic vapor phase epitaxy, J Appl Phys 113 (2013). https://doi.org/10.1063/1.4790383.

[41] A. Ashrafi, K. (Ken) Ostrikov, Raman-active wurtzite CdO nanophase and phonon signatures in CdO/ZnO heterostructures fabricated by nonequilibrium laser plasma ablation and stress control, Appl Phys Lett 98 (2011). https://doi.org/10.1063/1.3573795.

[42] A. Das, D. Singh, A. Kaur, C.P. Saini, D. Kanjilal, C. Balasubramanian, J. Ghosh, R. Ahuja, Temperature-Dependent Cationic Doping-Driven Phonon Dynamics Investigation in CdO Thin Films Using Raman Spectroscopy, The Journal of Physical Chemistry C 124 (2020) 21818–21828. https://doi.org/10.1021/acs.jpcc.0c06632.

[43] S.G. Choi, L.M. Gedvilas, S.Y. Hwang, T.J. Kim, Y.D. Kim, J. Zúñiga-Pérez, V. Muñoz Sanjosé, Temperature-dependent optical properties of epitaxial CdO thin films determined by spectroscopic ellipsometry and Raman scattering, J Appl Phys 113 (2013). https://doi.org/10.1063/1.4803876.

[44] A. Lysak, E. Przeździecka, A. Wierzbicka, R. Jakiela, Z. Khosravizadeh, M. Szot, A. Adhikari, A. Kozanecki, Temperature dependence of the bandgap of Eu doped {ZnCdO/ZnO}30 multilayer structures, Thin Solid Films 781 (2023) 139982. https://doi.org/10.1016/j.tsf.2023.139982.

[45] R. Bożek, Application of Kelvin Probe Microscopy for Nitride Heterostructures, Acta Phys Pol A 108 (2005) 541–554.

[46] R. Ferro, J.A. Rodríguez, Influence of F-doping on the transmittance and electron affinity of CdO thin films suitable for solar cells technology, Solar Energy Materials and Solar Cells 64 (2000) 363–370. https://doi.org/10.1016/S0927-0248(00)00228-2.

[47] E. Przeździecka, A. Lysak, A. Adhikari, M. Stachowicz, A. Wierzbicka, R. Jakiela, Z. Khosravizadeh, P. Sybilski, A. Kozanecki, Influence of the growth temperature and annealing on the optical properties of {CdO/ZnO}30 superlattices, J Lumin 269 (2024) 120481. https://doi.org/10.1016/j.jlumin.2024.120481.

[48] K.M. Yu, D.M. Detert, G. Chen, W. Zhu, C. Liu, S. Grankowska, L. Hsu, O.D. Dubon, W. Walukiewicz, Defects and properties of cadmium oxide based transparent conductors, J Appl Phys 119 (2016). https://doi.org/10.1063/1.4948236.

[49] Z. Zhang, J.T. Yates, Band Bending in Semiconductors: Chemical and Physical Consequences at Surfaces and Interfaces, Chem Rev 112 (2012) 5520–5551. https://doi.org/10.1021/cr3000626.

[50] B.J. Rodriguez, W.-C. Yang, R.J. Nemanich, A. Gruverman, Scanning probe investigation of surface charge and surface potential of GaN-based heterostructures, Appl Phys Lett 86 (2005). https://doi.org/10.1063/1.1869535.

[51] A. Dewasi, A. Mitra, Effect of variation of thickness of TiO2 on the photovoltaic performance of n-TiO2/p-Si heterostructure, Journal of Materials Science: Materials in Electronics 28 (2017) 18075–18084. https://doi.org/10.1007/s10854-017-7751-3.

[52] Ş. Karataş, A.A. Al-Ghamdi, F. Al-Hazmi, O.A. Al-Hartomy, F. El-Tantawy, F. Yakuphanoglu, The electrical properties of nanocluster-CdO/p-type silicon heterojunction structure at room temperature, Optoelectronics and Advanced Materials - Rapid Communications 6 (2012) 965–970.

[53] M. Sağlam, A. Ateş, M.A. Yıldırım, B. Güzeldir, A. Astam, Temperature dependent current–voltage characteristics of the Cd/CdO/n–Si/Au–Sb structure, Current Applied Physics 10 (2010) 513–520. https://doi.org/10.1016/j.cap.2009.07.011.

[54] D.A. Aldemir, Analysis of current–voltage and capacitance–voltage characteristics of Zr/p-Si Schottky diode with high series resistance, Modern Physics Letters B 34 (2020) 2050095. https://doi.org/10.1142/S0217984920500955.

[55] R.A. Ismail, A.-M.E. Al-Samarai, S.J. Mohmed, H.H. Ahmed, Characteristics of nanostructured CdO/Si heterojunction photodetector synthesized by CBD, Solid State Electron 82 (2013) 115–121. https://doi.org/10.1016/j.sse.2013.02.035.

[56] O.A. Hamadi, Characteristics of CdO—Si heterostructure produced by plasma-induced bonding technique, Proceedings of the Institution of Mechanical Engineers, Part L: Journal of Materials: Design and Applications 222 (2008) 65–72. https://doi.org/10.1243/14644207JMDA56.

[57] R.A. Ismail, O.A. Abdulrazaq, A new route for fabricating CdO/c-Si heterojunction solar cells, Solar Energy Materials and Solar Cells 91 (2007) 903–907. https://doi.org/10.1016/j.solmat.2007.02.006.

[58] I.L.P. Raj, N. Chidhambaram, S. Saravanakumar, S. Sasikumar, S. Varadharajaperumal, D. Alagarasan, T. Alshahrani, Mohd. Shkir, S. AlFaify, A comprehensive study on effect of annealing on structural, morphological and optical properties of CdO and photodetection of heterojunction n-CdO/p-Si diode, Optik (Stuttg) 241 (2021) 166406. https://doi.org/10.1016/j.ijleo.2021.166406.

[59] M. Soylu, H.S. Kader, Photodiode Based on CdO Thin Films as Electron Transport Layer, J Electron Mater 45 (2016) 5756–5763. https://doi.org/10.1007/s11664-016-4819-4.

[60] E. Płaczek-Popko, K.M. Paradowska, M.A. Pietrzyk, A. Kozanecki, Carrier transport mechanisms in the ZnO based heterojunctions grown by MBE, Opto-Electronics Review 25 (2017) 181–187. https://doi.org/10.1016/j.opelre.2017.06.010.

[61] E. Zielony, E. Płaczek-Popko, P. Nowakowski, Z. Gumienny, A. Suchocki, G. Karczewski, Electro-optical characterization of Ti/Au–ZnTe Schottky diodes with CdTe quantum dots, Mater Chem Phys 134 (2012) 821–828. https://doi.org/10.1016/j.matchemphys.2012.03.075.

[62] O. Breitenstein, P. Altermatt, K. Ramspeck, M.A. Green, J. Zhao, A. Schenk, Interpretation of the Commonly Observed I-V Characteristics of C-SI Cells Having Ideality Factor Larger Than Two, in: 2006 IEEE 4th World Conference on Photovoltaic Energy Conference, IEEE, 2006: pp. 879–884. https://doi.org/10.1109/WCPEC.2006.279597.

[63] Y. Huang, X. Gong, Y. Meng, Z. Wang, X. Chen, J. Li, D. Ji, Z. Wei, L. Li, W. Hu, Effectively modulating thermal activated charge transport in organic semiconductors by precise potential barrier engineering, Nat Commun 12 (2021) 21. https://doi.org/10.1038/s41467-020-20209-w.

[64] M. Rajini, S. Vinoth, K. Hariprasad, M. Karunakaran, K. Kasirajan, N. Chidhambaram, T. Ahamad, S.M. Alshehri, Tuning the optoelectronic properties of n-CdO:Fe/p-Si photodiodes fabricated by facile perfume atomizer technique for photo-detector applications, Applied Physics B 127 (2021) 109. https://doi.org/10.1007/s00340-021-07658-x.

[65] L.-H. Zeng, M.-Z. Wang, H. Hu, B. Nie, Y.-Q. Yu, C.-Y. Wu, L. Wang, J.-G. Hu, C. Xie, F.-X. Liang, L.-B. Luo, Monolayer Graphene/Germanium Schottky Junction As High-Performance Self-Driven Infrared Light Photodetector, ACS Appl Mater Interfaces 5 (2013) 9362–9366. https://doi.org/10.1021/am4026505.

[66] L. Zeng, S. Lin, Z. Lou, H. Yuan, H. Long, Y. Li, W. Lu, S.P. Lau, D. Wu, Y.H. Tsang, Ultrafast and sensitive photodetector based on a PtSe2/silicon nanowire array heterojunction with a multiband spectral response from 200 to 1550 nm, NPG Asia Mater 10 (2018) 352–362. https://doi.org/10.1038/s41427-018-0035-4.

[67] B. Nie, J. Hu, L. Luo, C. Xie, L. Zeng, P. Lv, F. Li, J. Jie, M. Feng, C. Wu, Y. Yu, S. Yu, Monolayer Graphene Film on ZnO Nanorod Array for High-Performance Schottky Junction Ultraviolet Photodetectors, Small 9 (2013) 2872–2879. https://doi.org/10.1002/smll.201203188.

[68] C. Lan, C. Li, S. Wang, T. He, T. Jiao, D. Wei, W. Jing, L. Li, Y. Liu, Zener Tunneling and Photoresponse of a $WS_2$/Si van der Waals Heterojunction, ACS Appl Mater Interfaces 8 (2016) 18375–18382. https://doi.org/10.1021/acsami.6b05109.

[69] M.B.A. Bashir, E.Y. Salih, A.H. Rajpar, G. Bahmanrokh, M.F. Mohd Sabri, The impact of laser energy on the photoresponsive characteristics of CdO/Si visible light photodetector, Journal of Micromechanics and Microengineering 32 (2022) 085006. https://doi.org/10.1088/1361-6439/ac7d93.